# An FPGA-based readout chip emulator for the CMS ETL detector upgrade

L. Zhang,[a] C. Edwards,[b] D. Gong,[b] X. Huang,[a] J. Lee,[c] C. Liu,[a] T. Liu,[a,1] T. Liu,[b]
J. Olsen,[b] Q. Sun,[b] J. Wu,[b] J. Ye,[a] and W Zhang[a]

[a] *Southern Methodist University,
Dallas, TX 75275, USA*

[b] *Fermi National Accelerator Laboratory,
Batavia, IL 60510, USA*

[c] *University of Illinois Chicago,
Chicago, Illinois 60607, USA*

 *E-mail:* `tliu@smu.edu`

ABSTRACT: We present an FPGA-based readout chip emulator board for the CMS Endcap Timing Layer (ETL) detector upgrade. The emulator board uses an Intel Cyclone 10 GX FPGA to emulate the digital functions of four Endcap Layer Readout Chips (ETROCs). Based on the actual ETROC design, the firmware is implemented and verified. The emulator board is being used for the ETROC digital design verification and system development.

---

[1] Corresponding author.

# Contents

## 1. Introduction

The MIP (Minimum Ionizing Particle) Timing Detector (MTD) is being developed to mitigate the pile-up effects after the CMS HL-LHC upgrade [1]. The MTD includes the Barrel Layer (BTL) and the Endcap-Timing Layer (ETL). In the ETL, Endcap Timing Read-Out Chips (ETROCs) are designed to read out Low-Gain Avalanche Detectors (LGADs) [2]. Each ETROC measures the Time Of Arrival (TOA) and the Time Over Threshold (TOT). The TOT is an estimation of the signal amplitude for the time-walk correction. Two prototypes, including ETROC0 [3] and ETROC1, have been taped out and evaluated. The third prototype ETROC2 [4] will be submitted in October 2022. The block diagram of ETROC2 is shown on the left side of figure 1. ETROC2 includes 16×16 pixels on the top and a global readout logic circuit on the bottom. Each pixel has a Pre-Amplifier (PA), a discriminator, a Time-to-Digital Converter (TDC) [5-6], and buffers. The threshold of the discriminator can be programmed with a Digital-to-Analog Converter (DAC) [7]. The global readout logic circuit consists of a Phase-Locked Loop (PLL) [8], a fast command decoder, a digital readout logic circuit, and an I$^2$C target module. Each ETROC2 receives a differential 40 MHz clock (CLK40) and a differential 320 Mbps Fast Command (FC). Each ETROC2 sends its data through its left and right differential digital outputs (DOL and DOR) at a data rate of 320 Mbps, 640 Mbps, or 1.28 Gbps.

The ETL readout system is shown on the right side of figure 1. Multiple (12, 24, or 28) ETROCs are connected to a readout board. Each readout board has a Versatile Link Plus Transceiver (VTRx+) [9] module, two Low Power Giga-Bit Transceiver (lpGBT) [10] chips, and a Giga-Bit Transceiver Slow-Control Adapter (GBT SCA) [11]. By using optical fibers, the VTRx+ module communicates with the off-detector Data Acquisition (DAQ) system. The first lpGBT recovers clocks and fast commands from a downstream link for the ETROCs, while both lpGBT chips transmit their data to the DAQ system through two upstream optical links. The GBT SCA provides slow control of all ETROCs. A power mezzanine board is mounted to the readout board to provide supply voltages.

To decouple the system-level development from the ASIC design, an FPGA-based ETROC emulator is developed. The emulator follows the same electrical interface, including Input/Output (I/O) pins, the fast command definition, and the output data frame definition, as ETROC2. In figure 1, an emulator board replaces the middle ETROC module. The emulator board was ready before February 2022, whereas the ETROC2 chips are expected to become available in February 2023. The emulator has been successfully used for the tests of the first readout board prototype, and the design guidelines for the second readout board prototype are provided. At the same time, by using the emulator, firmware and software have been developed for testing the readout board and the off-detector DAQ system. At least one year of design and test time has been saved. Moreover, a lot of testing preparation and debugging work has been done with the emulator before ETROC2 chips arrive.

The ETROC2 emulator has also helped us in the design verification of ETROC2, in particular, the system interfaces. Hardware emulation is far superior to pure simulation in terms of processing time. For example, it takes just one second to emulate the same amount of digital output data in the FPGA emulator as it would take weeks to simulate. The emulator approach has become a crucial part of the ETROC design verification methodology.

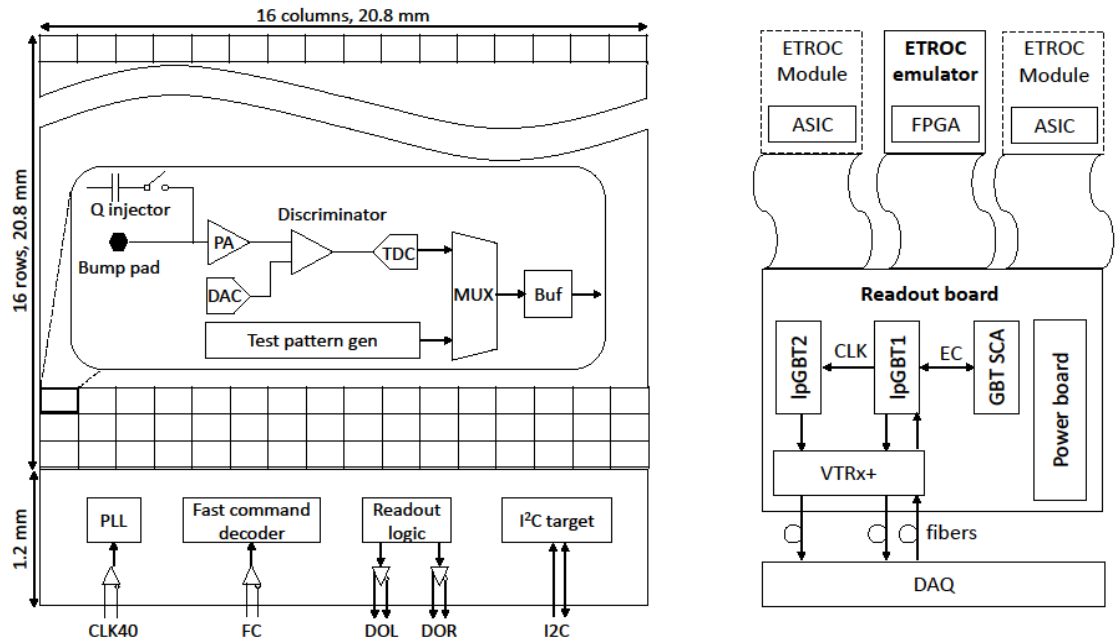

**Figure 1**. Block diagram of the ETROC2 (left) and the ETL readout system (right).

## 2. Hardware design

The block diagram of the emulator (left) and a photograph of an emulator board (right) are shown in figure 2. Each emulator board mimics the digital functions of four ETROC2 chips. To ease the firmware test, we emulate the counterparts (a command generator, a data checker, and an I$^2$C controller) that belong to the readout board/DAQ system in the same FPGA. These counterparts are marked as the DAQ emulator in figure 2. To generate four clock outputs, we use a clock fanout chip (Part No. CDCLVD1208 produced by Texas Instruments). The emulator has two cables to communicate with the readout board. The first cable (shown in blue in figure 2) is a 6-inch 34AWG Twinax cable with 14 differential pairs. Note that the cable has only two differential pairs of fast-command signals. Each differential pair of fast-command signals are multiply dropped to two I/O banks. The other cable (shown in gray in figure 1) is a 24-pin flat ribbon cable for power and slow control signals. The emulator requires a single 12 V supply voltage with a current of about 0.5 A. All the supply voltages needed on the board are generated through DC-DC converters or Low-Drop-Out (LDO) regulators. The emulator has passive loads for the readout board to monitor supply currents. The Printed Circuit Board (PCB) of the

– 2 –

emulator has 10 layers. The emulator is 165 mm long and 43 mm wide. The width of the emulator matches that of ETROC modules.

Considering the logic compatibility and the data rate range of general-purpose I/O pins, we select an Intel Cyclone 10 GX FPGA (Part No. 10CX220YF780E5G). This FPGA has five high-performance I/O banks that support differential 1.2 V Stub Series Terminated Logic (SSTL-12) logic and Low-Voltage Differential Signaling (LVDS). Among these high-performance I/O banks, four banks emulate the functions of four ETROC chips, and the other bank emulates the readout board/DAQ system. The FPGA has a high-range I/O bank that supports an input/output supply voltage (VCCIO) of 1.8 V or higher. This high-range I/O bank is used for a JTAG interface and flash memory. The emulator uses LVDS for the input signals CLK40 and FC. The output signals (DOL and DOR) and the I$^2$C signals have two options. The data output signals (DOL and DOR) are 1.2 V SSTL and DC coupled at 320 Mbps or 640 Mbps on the current emulator board. The I$^2$C signals are 1.2 V LVCMOS. In the other option, the data output signals (DOL and DOR) are LVDS and AC coupled, and the I$^2$C signals are 1.8 V LVCMOS with level shifters. A new emulator is under design to implement the latter option.

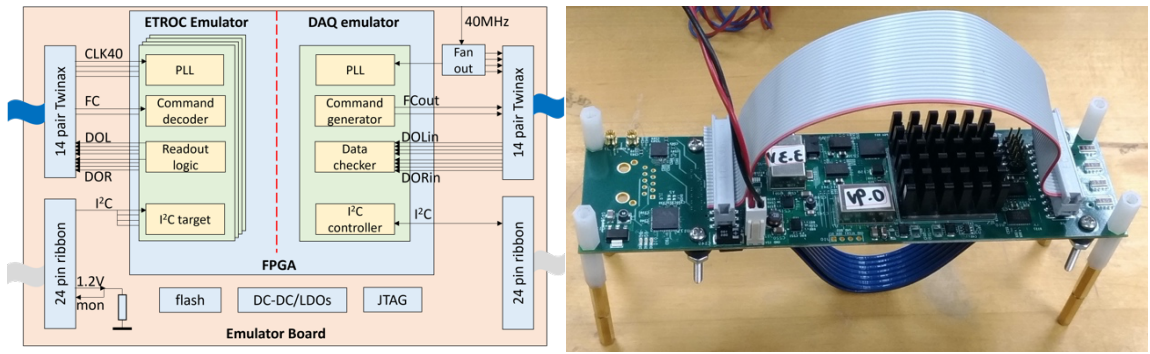

**Figure 2**. Block diagram of the emulator (left) and photograph of an Emulator board with cables plugged for loopback tests (right).

## 3. Firmware design

Based on the digital design of ETROC2, the firmware is being developed and verified on the emulator board. The I$^2$C target is relatively simple and will not be discussed in detail in this paper. The fast commands implemented in ETROC2 are verified first in the emulator. The data rate of the fast-command signal is 320 Mbps, and each fast command has eight bits. Fast commands are encoded in a Hamming code with a minimum distance of 3 bits between any two commands. Any single-bit flip, for example, a radiation-induced single event upset, can be detected or even corrected. The fast-command decoder searches for the command boundary and recovers commands from the serial fast-command bitstream.

The digital readout logic circuit is the core of firmware development. The data to be read out are either dummy TDC data or test patterns. The TDC data include a 10-bit TOA, a 9-bit TOT, a 10-bit Calibration code, and a Dada Valid (DV) bit. The DV bit indicates if the pixel is hit. The data of each pixel are first stored in a circular buffer. Once the command Level-1 Accept (L1A) is received, the corresponding data are transferred from the circular buffer to an L1A event buffer.

The trigger-selected data of all channels are transmitted from all L1A event buffers to a global data stream buffer via a switching network. The block diagram of the switching network [12] is shown in figure 3. Each pixel has an event buffer, a D flip-flop, and a switching cell. The switching cell is a combinational logic circuit that controls the order of the data transmission. Each switching cell has its data (marked as a red arrow) and two neighbors. The upstream neighbor is marked as a blue arrow and the downstream neighbor is marked as a green arrow. The data pass through the pixels with empty data and propagate downward, whereas the control signals propagate in the opposite direction. A pixel always transmits its data before its upstream neighbor and blocks the control signals from its downstream if it is not empty. Only one pixel is processed within a bunch crossing clock cycle. The 16 pixels in a column form a column chain. In each column chain, the pixel on the bottom has the highest priority to transmit its data, whereas the pixel on the top has the lowest priority. At the bottom of the columns, 15 extra switching cells form a row chain and connect the 16 columns to the global data stream buffer. In the row chain, the pixel on the left of the center has the highest priority and the pixel on the rightmost has the lowest priority.

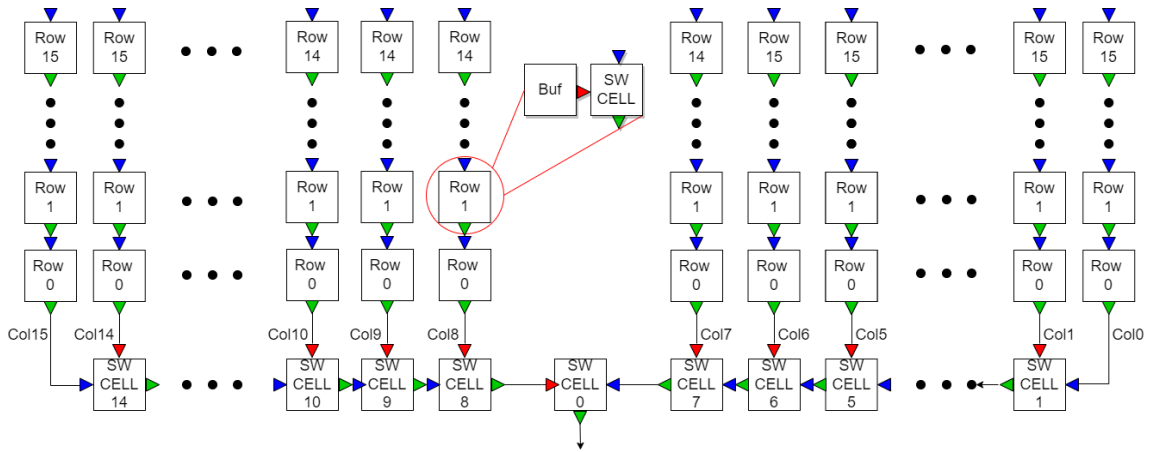

**Figure 3**. Block diagram of the switching network.

The data from the global data stream buffer are framed, scrambled, and serialized as a serial bitstream. After studying the performances of a few frame formats in the emulator, we implemented a frame format with a fixed length of 40 bits. The serial output data are the combination of four different frame types (headers, pixel data, trailers, and fillers). Every event has a header, data of multiple pixels that are hit, and a trailer. A frame of pixel data includes a 4-bit row address of the pixel, a 4-bit column address of the pixel, 10-bit TOA data, 9-bit TOT data, 10-bit calibration data, and three control/status bits. A trailer contains an eight-bit Cyclic Redundant Checking (CRC) code for frame integrity. If no data are transmitted, a filler is transmitted.

In addition to the pixel data path, ETROC2 implements a trigger path for monitoring purposes. A coarse map of user-defined hits is continuously sent out every bunch-crossing clock cycle through the trigger path. The granularity of trigger signals is programmable. The whole chip can be divided into 4x4, 2x2, 2x1, or 1x1 blocks to generate trigger data. The trigger path uses empty data in beam gaps to send out "flashing bits" for alignment. The trigger path and the pixel data path share the output bandwidth. The ratio of trigger data over pixel data is programmable. For example, if n ($0 \leq n < 7$) trigger bits are transmitted at 320 Mbps (DOL or DOR) every 25 ns, then 8-n bits can be used to transmit pixel data.

## 4. Hardware test and firmware verification

All of the digital circuits except the circular buffers and those implemented in the hardware Intelligent Property (IP) blocks (memory, serializer, deserializers, etc.) share the same source code as the ASIC and have been verified in the emulator. The verification was conducted on either a single emulator board with loopback cables or two connected emulator boards. In the latter setup, Emulator 1 received an external 40 MHz clock, fanned out the clock, and encoded fast commands. Emulator 2 generated test patterns and sent the data back to Emulator 1, based on the clock and fast command received from Emulator 1. Emulator 1 also checked the data generated in Emulator 2. The fast commands were monitored via a logic analyzer (Intel Signal Tap II) in both setups. Frames that violate frame generation rules (e.g., a header following another header) are considered frame format errors. The CRC code of each event was calculated and compared with the CRC value in the received trailer. Frame format and CRC errors were then counted and monitored. The system ran error-free for over 24 hours. Figure 4 is a screenshot of the data waveforms in the logic analyzer from a single emulator board with loopback cables. The signals shown on the screen are the left 1-bit trigger data, the left 40-bit pixel data, the left data valid bit, the right 1-bit trigger data, the right 40-bit pixel data, and the right data valid bit, respectively. All data shown in the screenshot are monitored on the DAQ emulator side after they are extracted from the raw bitstreams (DOL and DOR) at 320 Mbps. The 40-bit pixel data on the left side (DOL) contain a complete event (a header, two consecutive frames of pixel data, and a trailer). The 40-bit pixel data on the right side (DOR) include a complete event (a header, a pixel-data frame, a trailer) and a filler. The complete data events are marked in green and the filler is marked in blue.

A few signals were brought out of the FPGA to an oscilloscope to monitor the emulator operation. The data were received over three hours without any CRC or frame format errors. The I$^2$C target module of an emulator board was verified via a commercial USB I$^2$C dongle (Module USB-ISS from Robot-Electronics).

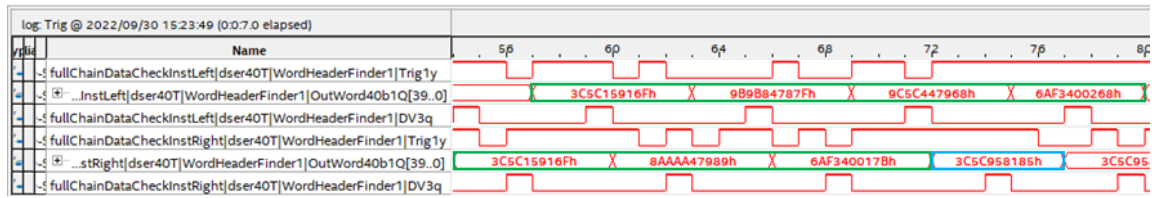

**Figure 4**. Screenshot of the data waveforms monitored in the logic analyzer.

The emulator is being used in the system testing of the readout board. In a test, the emulator board generated and transmitted a pseudo-random binary sequence (pattern $2^7$-1) on each digital output, while the readout board checked its received input data. The test was performed for over two weeks without any errors. The bit error rate for each data channel was estimated to be less than $8.9\times10^{-15}$ at a confidence level of 99%. In another test, a readout board communicated with three emulator boards and an FPGA evaluation board (Xilinx KCU105), which serves as the DAQ system. The data generated by the emulator boards can be extracted successfully.

## 5. Conclusion

An FPGA-based emulator has been developed for the CMS ETL detector upgrade. The emulator board has an Intel Cyclone 10 GX FPGA to emulate the functions of four ETROCs. Based on the actual ETROC design, the firmware is implemented and preliminarily verified. The emulator board is being used for the ETROC digital design verification and system development.

## Acknowledgments

The authors would like to thank Mr. Eric Hazen, Mr. Sergey V Los, Mr. Andrew Peck, Dr. Daniel Spitzbart, Prof. Indara Suarez, and Mr. Shouxiang Wu for useful discussions on the emulator interface with the readout board.